\documentclass[a4paper]{elsarticle}
\usepackage{graphicx}
\usepackage{amsmath}
\usepackage{amssymb}
\newcommand{\ia}{\'{\i}} 

\title{
Characterization of invariant patterns in a slowly rotated granular tumbler
}
\author[usb]{Leonardo Reyes\corref{1}}
\author[usb]{Oscar P\'erez}
\author[usb]{Claudia Colonnello}
\author[usb,ivic]{Ang\'elica Goncalves}
\author[usb]{Haydn Barros}
\author[ivic]{Iv\'an S\'anchez}
\author[usb]{Gustavo Guti\'errez}

\cortext[1]{Corresponding author. Tel.:+58 212 9063541. Email adress: lireyes@usb.ve}

\address[usb]{Departamento de F\ia sica, Universidad Sim\'on Bol\'ivar, Apartado Postal 89000, Caracas 1080-A, Venezuela}
\address[ivic]{Centro de F\ia sica, Instituto Venezolano de Investigaciones Cient\ia ficas, Caracas 1020-A, Venezuela}

\begin{document}

\begin{abstract}
We report experimental results of the pattern developed by a mixture of two types of grains in a triangular rotating tumbler operating in the avalanche regime. 
At the centroid of the triangular tumbler an invariant zone appears where the grains do not move relative to the tumbler.
We characterize this invariant zone by its normalized area, $A_i$, and its circularity index
as a function of the normalized filling area $A$. We find a critical filling area so that only
for $A>A_c$ invariant zones are obtained. These zones scale as $A_i\sim (A-A_c)^2$ near $A_c$. We have obtained a maximum in the circularity index for $A\approx 0.8$, for which the shape of the invariant zone is closer to a circular one.  The experimental results are reproduced by a simple model which, based on the surface position, accounts for all the possible straight lines within the triangle that satisfy the condition of constant $A$.
We have obtained an analytic expression for the contour of the invariant zone.
\end{abstract}

\maketitle

\begin{figure}
\centering
\includegraphics[clip,width=10cm]{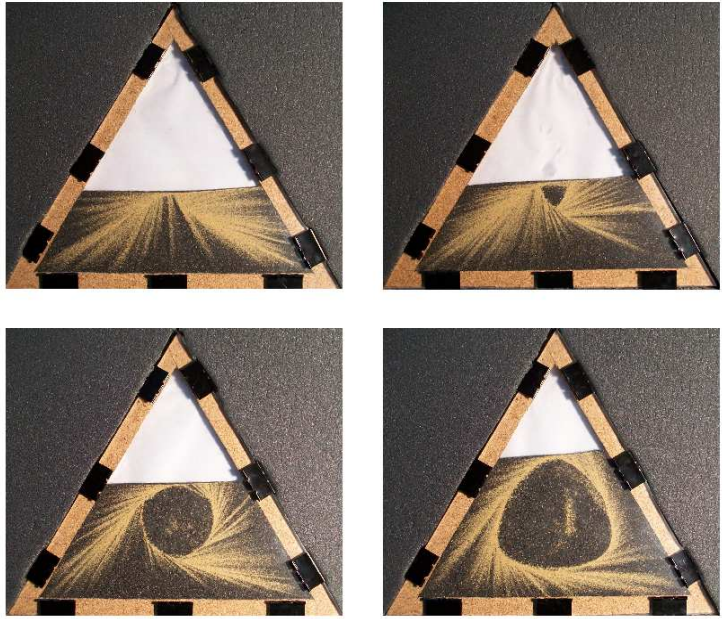}
\caption{The tumbler filled to four differents heights. No invariant zone appears for small filling heights. 
The axis of rotation is perpendicular to the page and this pictures were taken after three revolutions (see fig \ref{numV}).}
\label{imagenes}
\end{figure}

The physics of granular matter is an active field of research, the static and dynamic collective properties of a large number of grains are still 
under scrutiny \cite{deGennesRMP,ottino2006granular}. A rotating tumbler filled with grains is an interesting model system: it has direct 
industrial analogs \cite{boateng2011rotary,Grajales2012167,watkinson1982limestone} and
is simple enough for theory and numerical simulations \cite{hillEtAl99,Chand20124590,PhysRevE.60.1975}. Several regimes can be found in this type of 
system \cite{henein1983,rajFlow1990}, 
the Froude number $F_r=\omega^2L/g$
being one of the important parameters to characterize the experimental situation. Here $\omega$ is the angular speed of rotation, $L$ is a 
characteristic length of the tumbler and $g$ is the acceleration of gravity. For small $F_r$ we are in an avalanche regime, where grain flow occurs during localized episodes of short duration compared to the period of rotation. By increasing $\omega$ we 
can go to a continuous flow regime.
In the avalanche regime we typically find segregation, kinks and pattern formation. Two main types of segregation are usually observed: axial segregation in 
the direction parallel to the axis of rotation \cite{Taberlet2006,PhysRevE.56.4386,Alexander_2004,Kuo2005196},  and radial segregation in the
direction normal to the axis of rotation \cite{hsiau1,hsiau2,Khakhar2001232,CPLX:CPLX20082,ikerMullin2006,springerlink:10.1007/s10035-005-0198-x}. A combination of 
the two types can lead to complex patterns in special cases \cite{kuo2006}. When the rotating tumbler is thin enough, axial segregation is frustrated and only radial segregation is observed. In this article we report experimental results
for a rotating triangular tumbler filled with grains in the avalanche regime. A triangular tumbler has been reported to give very high mixing 
efficiencies \cite{McCarthy96}. We characterize the invariant zone that appears around the centroid of the triangle, in which the grains do not move relative to the tumbler \cite{McCarthy96,NatMetcalfe95}. 
The experimental results are reproduced by a simple model which can be solved {\it analytically} for the profile of the invariant zone which, to the best of our knowledge, has not been achieved before. This is relevant for the characterization of this pattern forming system in the small $F_r$ regime.

\begin{figure}
\centering
\includegraphics[clip,width=12cm]{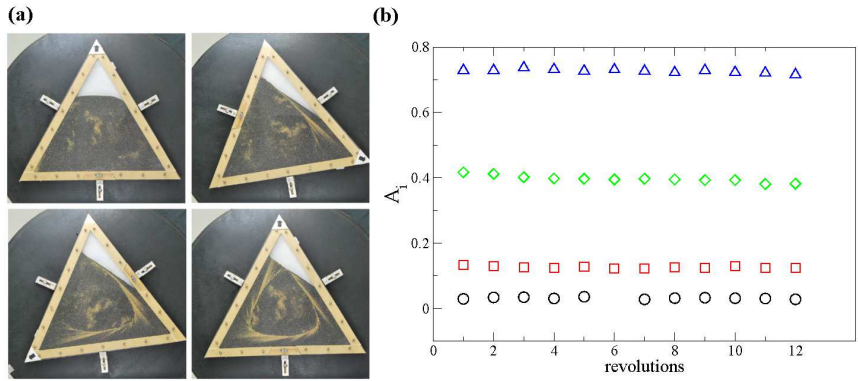}	
\caption{a) Initial condition and evolution of the system up to the first revolution. Note the arrow in one of the vertex of the triangle. 
For the preparation process the container is held with the axial axis vertically oriented and a few up and down taps with some horizontal component are manually produced until two layers of grains (the dark large grains at the top) well separated along the axial axis are easily observed.
b) Invariant area $A_i$ as a function of the number of revolutions for four different fillings areas, when the tumbler is rotated with the motor at $1.5rpm$.}
\label{numV}
\end{figure}

Our system consists of two kinds of grains, clear sand from the dunes of Coro, Venezuela, with particle sizes between $215-250\mu m$, and dark grey volcanic
sand from Puc\'on, Chile, with particle sizes between $350-600 \mu m$.  The tumbler is an equilateral triangle with side $L=30.5cm$. The grains are held
within two Plexiglas plates $7mm$ apart. 
The preparation of the system is done in the following way: a) The triangular container has a small opening in its side that allows for the introduction of the grains in the appropriate proportion; b) The small opening closes seamlessly and then is shaken horizontally so that the clear smoother and smaller particles percolate to the bottom plate originating a stratified granular medium with a thin layer of dark bigger and rougher particles and another thin layer of  clear particles; c) The container is carefully mounted in the support frame so that the two types of grains are initially segregated as much as posible forming two layers along the rotation axis; no radial segregation is present in our initial condition (see fig. \ref{numV}); d) 
Finally the triangular container is rotated at a constant frequency of $1.5rpm$ ($F_r\approx 5\times 10^{-3}$). The same results are obtained if the drum is rotated manually at a very slow rotation, demonstrating the generality of this process if the frequency is kept sufficiently low.
If the normalized filling area $A$ is large enough we observe an {\it invariant} zone as shown in figure 
\ref{imagenes}; this zone was called a {\it core} in ref. \cite{McCarthy96}.  In figure 
\ref{numV} we show the initial evolution of the system and the normalized size of the invariant zone $A_i$ as a function of the number of revolutions for four different filling areas. As with $A$, the invariant area $A_i$ was normalized with the total area of the triangle. Grains in this region are never in the avalanche region, and then, they never move with respect to each other or the drum \cite{Christov}.
Outside this invariant zone the grains segregate forming dark and clear bands \cite{makseNature97,PRL1997MakseStanley,maksePRL99} (see figure \ref{imagenes}).
We will show that, in fact, the development of the invariant zone is independent of the granular sample utilized (mono or polydisperse). Nevertheless, the two types of grains used in this experiment allow for a clearer visualization of the process.
 As reported in reference \cite{McCarthy96}, we observe that if we continue rotating the tumbler this invariant zone rotates very slowly relative to the tumbler without appreciably changing its size and shape, but we do not characterize this behaviour here. 

\begin{figure}
\centering
\includegraphics[clip,width=8cm]{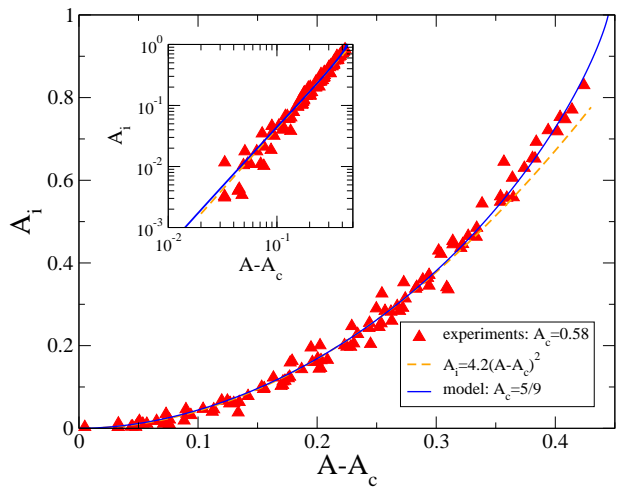}	
\caption
{Normalized invariant area $A_i$ as a function of $A-A_c$, where $A$ is the normalized filling area
and $A_c$ the critical value of $A$ below  which there is no invariant zone. Continuous line: geometrical model, with $A_c=5/9\approx 0.56$.
Triangles: experimental results with $A_c=0.02+5/9\approx 0.58$.
In the inset it is shown the data in a log-log scale. The relation $A_i\sim(A-A_c)^2$ (dashed line) describes well our results, up to $A-A_c\approx 0.35$. Within experimental error, we can not diferentiate between the manual way of rotation or the rotation with a motor. The difference in $A_c$ between the model and experiments ($\Delta A_c\approx 0.02$) can be interpreted by assuming a finite number of layers of grains involved in the avalanche zone (see text).}
\label{Ainvariante}
\end{figure}

In figure \ref{Ainvariante} we can see the normalized size of the invariant zone $A_i$ as a function of $A-A_c$: we find a critical filling area so that only
for $A>A_c$ invariant zones are obtained. Figure \ref{Ainvariante} was obtained by analyzing the 
digital images obtained from the experiments, and from the geometrical model described below. For the geometrical model $A_c=5/9\approx 0.56$, and for
the experiments $A_c=0.02+5/9\approx 0.58$. Appart from a slight displacent of $A_c$, $\Delta A_c\approx 0.02$, we see that the geometrical model describes well our experimental data. As can be seen in figure \ref{Ainvariante} the relation $A_i\sim(A-A_c)^2$ describes well our results for $A$ close to $A_c$. $\Delta A_c$ was calculated through a second order fitting procedure performed to the experimental data.

\begin{figure}
\centering
\includegraphics[clip,width=9cm]{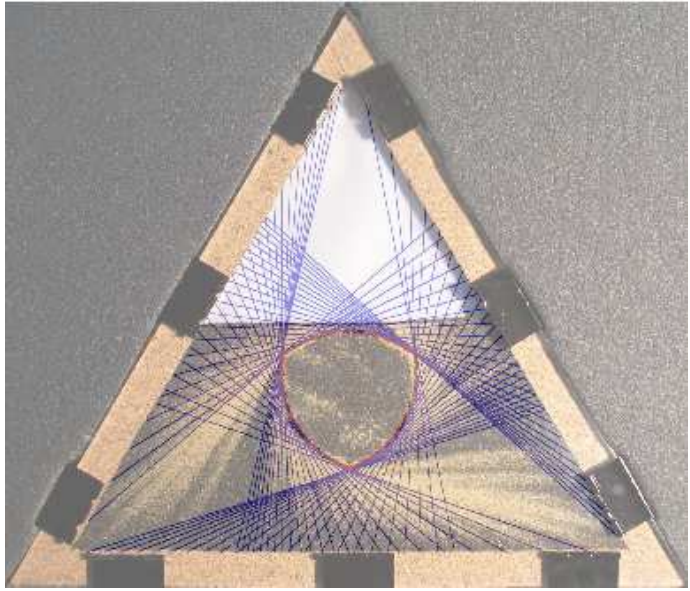}
\caption{Invariant pattern, experiment {\it versus} geometrical model. For the experiment the filling area is $A=0.75$. For the model $A=0.71$. Thus, for this  realization we used $\Delta A_c\approx 0.04$ (see text). The region around the centroid not intersected by any of the lines is the invariant zone. The contour with dashed lines are the profile given by equation (\ref{profile}) and rotations of $\pm 120^o$.}
\label{comparacion}
\end{figure}

Since we are in an avalanche regime, as we rotate the tumbler all the changes in the granular collective occur
in a narrow layer of grains close to its free surface. Under these conditions one can obtain the shape 
of the invariant zone from the following simple geometrical model. The free surface can be thought of as a line that intersects the triangle 
creating a partition into two regions, a filled region and an empty region. If the packing fraction of the bulk is constant, the area of these 
regions must be preserved by the rotation of the tumbler, which is equivalent to a rotation of the free surface line. 
To obtain the invariant zone for a given filling area $A$, we construct all the lines that intersect the triangle, preserving the area of the 
filled and empty regions. A similar model was proposed in reference \cite{McCarthy96} but no characterization of the invariant patterns was presented. In figure \ref{comparacion} we compare the geometrical model with an experimental image. The simplest way to solve the geometrical model is to consider the area of the empty region which is itself triangular. 

We can obtain the intercept $C$ along the $y$ axis of these area preserving lines as a function of their slope $m$ and the normalized area of the empty region $A_v$, with $A+A_v=1$. With the origin of coordinates in the centroid of the triangle, we obtain:

\begin{equation}{\label{eq1}
C(m)=\frac{1}{\sqrt{3}}{\frac{1+B_m-3\sqrt{A_v B_m}}{1+B_m}}}
\end{equation}
for a given $A_v$, with
\begin{equation}{\label{eq2}
B_m=\frac{\sqrt{3}+m}{\sqrt{3}-m}.}
\end{equation}
From equation (\ref{eq1}) and an inverse Legendre transform \cite{Callen} we can obtain the profile:
\begin{equation}{\label{profile}
y(x)=\frac{1}{\sqrt{3}}\left( 1-\frac{3}{2} \sqrt{4x^2+A_v}  \right).}
\end{equation}
The profile (\ref{profile}) can be seen in figure \ref{comparacion}, which goes from $-x_f$ to $x_f$, with $x_f=\frac{3\sqrt{1-2A_v}-1}{8}$. The rest of the contour of the invariant zone are rotations of $\pm 120^o$ of (\ref{profile}). From profile (\ref{profile}) we can obtain the area $A_i$ and perimeter $P_i$ of the invariant zone: $\frac{\sqrt{3}}{4}A_i= 6\int_0^{x_f}{ydx-\sqrt{3}x_f^2}$ and $P_i=6\int_0^{x_f}{\sqrt{1+y'^2}dx}$.

One finds a special value of the filling area $A_c=5/9\approx0.56$, such that the profile given by equation (\ref{profile}) satisfies $y(0)>0$ only for $A>A_c$. Above $A_c$ there is an invariant zone, and below $A_c$ the whole surface of the 
triangle is crossed by free surface lines.  The displacement in $A_c$ obtained experimentally, $\Delta A_c\approx 0.02$, can thus be explained in terms of a finite width of the avalanche region. In particular, if we need $n$ layers of grains above the centroid in order to obtain an invariant zone, and with $d/L\approx 10^{-3}$, we get $n\approx\frac{3\sqrt{3}}{8}\Delta A_c \frac{L}{d}\approx 13$, which seems reasonable since it represents a moving layer of around $4mm$ which is in agreement with direct observations during the avalanche processes. See reference \cite{Khakhar2008} for a detailed study of the thickness of the flowing layer in tumblers.

\begin{figure}
\centering
\includegraphics[clip,width=8cm]{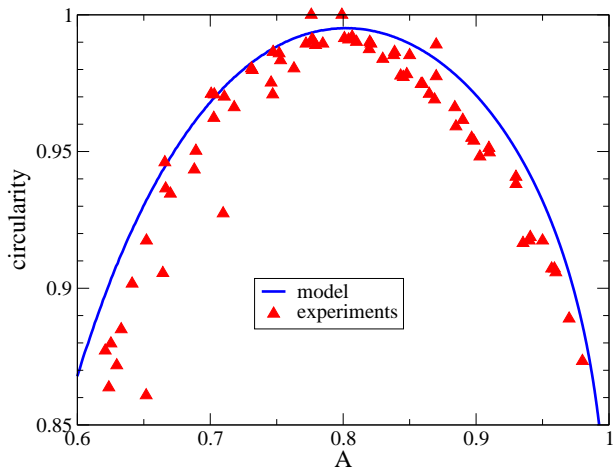}	
\caption
{Circularity index vs. normalized filling area $A$. Triangles: experimental results; circles: geometrical model. 
 The model reproduces fairly
well the experimental data. For $A\approx 0.8$ the invariant area is closer in shape to a circle. Since no displacement in $A$ was used for this figure, $\Delta A_c$ used in figure \ref{Ainvariante} in only needed for the {\it size} of the invariant zone.}
\label{circularidad}
\end{figure}

In order to characterize the shape of the invariant zones we calculated the circularity index of the figures obtained from the experiments and from 
the geometrical model. This index is defined as the ratio of the perimeter of a circle with the same area of the invariant zone to the perimeter of the invariant zone. This index is shown in figure \ref{circularidad}. We obtain a transition from an inverted small triangle to an up-right large triangle (see 
also fig \ref{imagenes}),
the transition point being at $A\approx 0.8$ in which the shape of the invariant zone is closer to the shape of a circle (a circle has a circularity index of $1$). Again,
we observe a good agreement between the experiments and the model. Note that no displacement in $A$ in needed in this case, so $\Delta A_c$ is needed only for the size of the invariant zone, not its shape.
 
In summary, in this article we have characterized the invariant zone that develops in a slowly rotated triangular tumbler. 
This invariant zone appears for filling areas
$A$ greater than a critical filling area $A_c$. 
Our results are reproduced with a simple geometrical model that considers all lines within the triangle such that the filling area $A$ is kept constant. The size 
of the invariant zone scales as $A_i\sim (A-A_c)^2$ near $A_c$; we have found a displacement in $A_c$ between the model and the experiments which can be interpreted in terms of avalanches involving a finite number of layers of grains. This displacement is needed only to correct the size of the invariant zone, not its shape.
We have obtained a maximum in the circularity index for $A\approx 0.8$, for which the
shape of the invariant zone is closer to a circular one. 
It is interesting that this complex system, a tumbler operated in the avalanche regime, 
can generate patterns that can be understood within geometry alone.
An interesting future work would be to study, experimentally and within the geometrical model, the {\it erosion} of the invariant zone that occurs if we 
increase the angular speed of rotation of the tumbler. 

We would like to thank Armando Uruburu and Augusto Vito for technical assistance.


\bibliographystyle{unsrt}

\end{document}